\documentclass[twocolumn,aps,showpacs,preprintnumbers,amsmath,amssymb]{revtex4}

\usepackage{graphicx}
\usepackage{dcolumn}
\usepackage{bm}

\begin{document}

\title{Enhancement of fusion at near and sub-barrier energies 
for neutron-rich light nuclei}

\author{Varinderjit Singh}
\author{J. Vadas}
\author{T.~K. Steinbach}
\author{B.~B. Wiggins}
\author{S. Hudan}
\author{R.~T. deSouza}
\email{desouza@indiana.edu}
\affiliation{%
Department of Chemistry and Center for Exploration of Energy and Matter, Indiana University\\
2401 Milo B. Sampson Lane, Bloomington, Indiana 47408 USA}%

\author{Zidu Lin}
\author{C.~J. Horowitz}
\affiliation{%
Department of Physics and Center for Exploration of Energy and Matter, Indiana University\\
2401 Milo B. Sampson Lane, Bloomington, Indiana 47408 USA}%

\author{L.~T. Baby}
\author{S.~A. Kuvin}
\author{Vandana Tripathi}
\author{I. Wiedenh\"{o}ver}
\affiliation{
Department of Physics, Florida State University, Tallahassee, Florida, 32306 USA}%

\date{\today}

\begin{abstract}
\begin{description}
\item[Background] Measurement of the fusion cross-section for neutron-rich light nuclei is crucial in ascertaining if fusion of these nuclei occurs in the outer crust of a neutron star.
\item[Purpose] To measure the fusion excitation function at near-barrier energies for the $^{19}$O + $^{12}$C system 
and compare the experimental results with the fusion excitation function of
$^{18}$O + $^{12}$C and $^{16}$O + $^{12}$C.
\item[Method] A beam of $^{19}$O, produced via the $^{18}$O(d,p) reaction, was incident on a $^{12}$C target at energies near the Coulomb barrier. Evaporation residues 
produced in fusion of $^{18,19}$O ions with $^{12}$C target nuclei were detected with good geometric efficiency and identified by measuring their
energy and time-of-flight. 
\item[Results] A significant enhancement 
in the fusion probability of $^{19}$O ions with a $^{12}$C target as compared to $^{18}$O ions is observed. 
\item[Conclusion] The significantly larger cross-sections observed at near barrier energies are not 
predicted by a static model of fusion for $^{19}$O + $^{12}$C indicating that dynamics play an important
role in the fusion of neutron-rich light nuclei. 
\end{description}
\end{abstract}

\pacs{26.60.Gj, 25.60.Pj, 25.70.Jj}
\maketitle

The conversion of matter to energy in the fusion of two nuclei is an extremely important process. Not only is it
responsible for element production whether in stellar interiors \cite{Burbridge57} or terrestially 
\cite{Khuyagbaatar14, Yanez14}, but it is also  
associated with the release of a tremendous amount of energy \cite{Oliphant35}. 
It has recently been proposed that fusion reactions are responsible for
fueling the energy release associated with X-ray superbursts \cite{Horowitz08}. In an X-ray superburst, an accreting neutron
star releases as much energy (10$^{42}$ ergs) in a few hours 
as our sun does in an entire decade 
\cite{Cumming01}. 
The fusion reaction proposed to fuel the X-ray superburst is fusion of $^{12}$C nuclei. However, the temperature of the 
neutron star crust appears to be too low for this reaction to occur. It has been hypothesized that fusion of two
neutron-rich light nuclei 
occurs more readily than the corresponding $\beta$-stable isotopes, providing 
a heat source that triggers the $^{12}$C fusion \cite{Horowitz08}.
While the fusion of both light and heavy $\beta$-stable nuclei has been studied for 
several decades, only 
recently has the investigation of fusion of neutron-rich nuclei become feasible due to 
radioactive beam facilities \cite{Liang03, Loveland06, Lemasson09, Kolata12, Kohley13}. 

Fusion of two nuclei is impacted both by the structure of the two nuclei as well as the dynamics of the fusion process. 
While for heavy nuclei the role of collective modes during the fusion process is well established \cite{Wu86}, the
fusion of light nuclei near $\beta$-stability is reasonably well described without inclusion of significant fusion dynamics.
Investigating fusion using an isotopic chain of projectile nuclei provides a unique opportunity.  
As the charge distribution of the projectile nuclei is essentially 
unaffected by the additional neutrons, the repulsive Coulomb potential is largely unchanged. Consequently, the 
comparison of the fusion excitation functions 
for the different projectiles provides access to the change in the attractive nuclear potential. This change in the attractive potential 
can be related to changes in the neutron density distribution with increasing number of neutrons.

An initial measurement of fusion induced with neutron-rich oxygen nuclei suggested 
an enhancement of the fusion probability as compared to a standard fusion-evaporation model 
\cite{Rudolph12}. 
Recently, the fusion of $^{10,14,15}$C + $^{12}$C 
has been investigated using a novel active target 
approach \cite{Carnelli14}. At the above barrier energies 
studied, no significant fusion enhancement was observed relative to a simple barrier penetrability model. Close examination of the $^{15}$C + $^{12}$C data presented reveals that while the data manifest a lower value of the astrophysical S factor at higher energies as compared to the model, at lower energies there is an indication that the data exceed the model predictions. 
It can be argued on general grounds that fusion enhancement is best studied at near and 
below barrier energies.
At low incident energies, the importance 
of central (low l-wave) collisions, which emphasize the attractive nuclear interaction, is enhanced. 
In addition, low relative energy allows the 
changing internuclear potential to effectively couple with
collective excitations in the two nuclei, resulting in a fusion enhancement.

To definitively establish if neutron-rich light nuclei exhibit a fusion 
enhancement at sub-barrier energies, high quality 
experimental data is needed. As beams of neutron-rich nuclei far from $\beta$-stability, 
which will provide the most stringent test of fusion
models, will be available at low intensity, it is crucial to develop an experimental 
approach capable of measuring the fusion
cross-section using such low-intensity radioactive beams. 
In this work, we present for the first time a measurement of the 
total fusion cross-section for $^{19}$O + $^{12}$C at incident energies near the barrier 
and compare the results with the fusion cross-section for $^{16,18}$O + $^{12}$C as well as a theoretical model of fusion.

\begin{figure}
\includegraphics[scale=0.7]{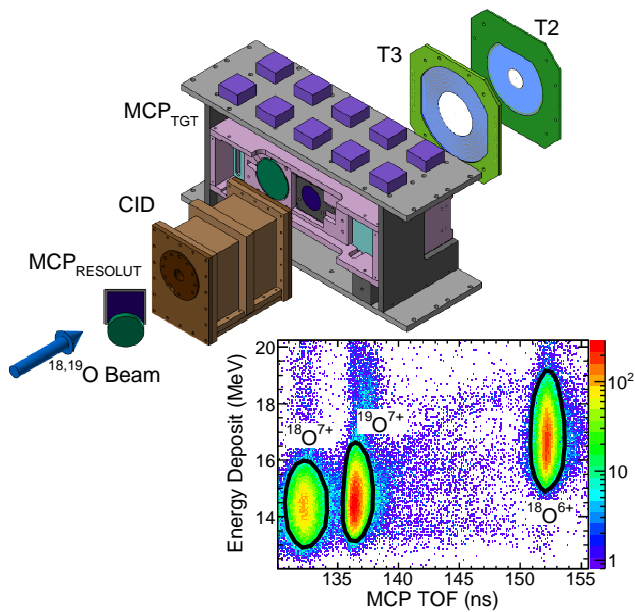}
\caption{\label{fig:setup} (Color online) Schematic illustration of the experimental setup. 
The MCP$\mathrm{_{RESOLUT}}$ detector is located approximately 3.5 m upstream of 
the compact ionization chamber (CID) which is situated directly in front of the 
MCP$\mathrm{_{TGT}}$ detector.
Inset: Energy deposit versus time-of-flight spectrum for ions exiting RESOLUT that are incident on $^{12}$C target 
at E$_{\mathrm{lab}}$=46.7 MeV.
Color is used to represent yield in the two-dimensional spectrum.}
\end{figure}

The experiment was performed at the John D. Fox accelerator laboratory at Florida State University. A beam of $^{18}$O ions, accelerated to an 
energy of 80.7 MeV, impinged on a deuterium gas cell at a pressure of 350 torr and cooled to a temperature of 77 K. Ions of $^{19}$O were produced via 
a (d,p) reaction and separated from the incident beam by the electromagnetic spectrometer RESOLUT \cite{RESOLUT}. 
Although this spectrometer rejected 
most of the unreacted beam that exited the production gas cell, the beam exiting the spectrometer 
consisted of both $^{19}$O and $^{18}$O ions necessitating identification of each ion on a 
particle-by-particle basis. 
Identification of each incident beam particle in the beam mixture allowed simultaneous measurement of $^{18}$O + $^{12}$C 
and $^{19}$O + $^{12}$C. This dual measurement provided a robust measure of the potential fusion enhancement.
The setup used to measure fusion of oxygen ions with carbon nuclei in this experiment 
is depicted in Fig.~\ref{fig:setup}. 
To identify beam particles, the energy deposit ($\Delta$E) and time-of-flight (TOF) 
of each particle was measured prior to the target. 
After exiting the RESOLUT spectrometer particles traverse a thin foil (0.5 $\mu$m thick aluminized mylar) ejecting electrons in the process. 
These electrons are accelerated and bent out of the beam path and onto the surface of a 
microchannel plate detector (MCP$\mathrm{_{RESOLUT}}$) where they are 
amplified to produce a fast timing signal. After traversing the thin foil of MCP$\mathrm{_{RESOLUT}}$, 
the oxygen ions passed through a compact 
ionization detector (CID) located approximately 3.5 m downstream
of MCP$\mathrm{_{RESOLUT}}$. 
In passing through this ionization chamber, ions deposit an energy
($\Delta$E) characterized 
by their atomic number (Z), mass number (A), and incident energy. 
After exiting CID the ions are incident on 
a 105 $\mu$g/cm$^2$ carbon foil. This carbon foil serves both as a secondary electron emission foil 
for the target microchannel plate detector (MCP$\mathrm{_{TGT}}$) and as the target for the 
fusion experiment \cite{Steinbach14}. Periodic insertion of a surface barrier, silicon detector directly into the beam path just prior 
to the target provided a measurement of the energy distribution of $^{19}$O and $^{18}$O ions incident on the target.

By utilizing the timing signals from both microchannel plate detectors together with the
energy deposit in the 
ionization chamber, 
a $\Delta$E-TOF measurement is performed. 
This measurement allows identification of ions in the beam as indicated in the inset of Fig.~\ref{fig:setup}. Clearly evident in the figure are three peaks associated with the $^{19}$O$^{7+}$ ions, $^{18}$O$^{7+}$ ions, and $^{18}$O$^{6+}$ ions. 
The cleanly identified $^{19}$O ions corresponded to approximately 31 \% of the beam intensity with 
the $^{18}$O$^{7+}$ and $^{18}$O$^{6+}$ corresponding 
to approximately 20 \% and 29 \% respectively. The intensity of the $^{19}$O beam incident on the target was 
1.5 - 4 x10$^3$ ions/s. 
Fusion of a $^{19}$O (or $^{18}$O) nucleus in the beam together with a $^{12}$C nucleus in the target 
foil results in the production of an excited $^{31}$Si 
(or correspondingly $^{30}$Si) nucleus. For collisions near the Coulomb barrier the excitation of the fusion product is relatively modest, E$^*$ $\approx$ 35 MeV. 
This fusion product de-excites by evaporation of a few neutrons, protons, and $\alpha$ particles resulting in an evaporation residue (ER). 
Statistical model calculations \cite{evapor} indicate that for a $^{31}$Si compound nucleus, the nuclei
$^{30}$Si, $^{29}$Si, $^{28}$Si, $^{29}$Al, $^{28}$Al, $^{27}$Mg, and $^{26}$Mg account for the bulk of the ERs. 
Emission of the light particles deflects the ER from the beam direction due to momentum conservation. 
The ERs are detected and identified by two annular silicon detectors designated T2 and T3 
situated downstream of the MCP$\mathrm{_{TGT}}$. These detectors  subtend the angular range 
3.5$^\circ$ $<$ $\theta_{lab}$ $<$ 25$^\circ$. Evaporation residues 
are distinguished from scattered beam, as well as emitted light particles, by measuring their time-of-flight 
between the MCP$\mathrm{_{TGT}}$ detector and the silicon detectors \cite{deSouza11} together with 
the energy deposit in the Si detector. Using the measured energy deposit and the time-of-flight, the mass of the ion can be calculated. By this means, ERs are cleanly distinguished from the incident beam \cite{Steinbach14a}.

The fusion cross-section is extracted from the measured yield of evaporation residues through the 
relation $\sigma_{fusion}$ = N$_{ER}$/($\epsilon_{ER}$ x t x N$_{I}$) where 
N$_{I}$ is the number of beam particles of a given type incident on the target, t is the target 
thickness, $\epsilon_{ER}$ is the detection efficiency, and N$_{ER}$ 
is the number of evaporation residues detected. The number N$_{I}$ is determined by counting the 
particles with the appropriate time-of-flight 
between the two microchannel plates that additionally have the correct identification in the $\Delta$E-TOF map 
depicted in the inset of Fig.~\ref{fig:setup}. 
The target thickness, t, of 105 $\mu$g/cm$^2$ is provided by the manufacturer and has an uncertainty 
of $\pm$ 0.5 $\mu$g/cm$^2$. The number of detected residues, 
N$_{ER}$, is determined by summing the number of detected residues clearly identified by the ETOF technique 
\cite{Steinbach14a}. 
To obtain the detection efficiency, $\epsilon_{ER}$,
a statistical model is 
used to describe the de-excitation of the fusion product together with the geometric acceptance of the experimental setup. 
The detection efficiency varied from 37 \% at the highest incident energies measured to 42 \% at the 
lowest incident energy due to the changing kinematics of the reaction.

\begin{figure}
\includegraphics[scale=1.0]{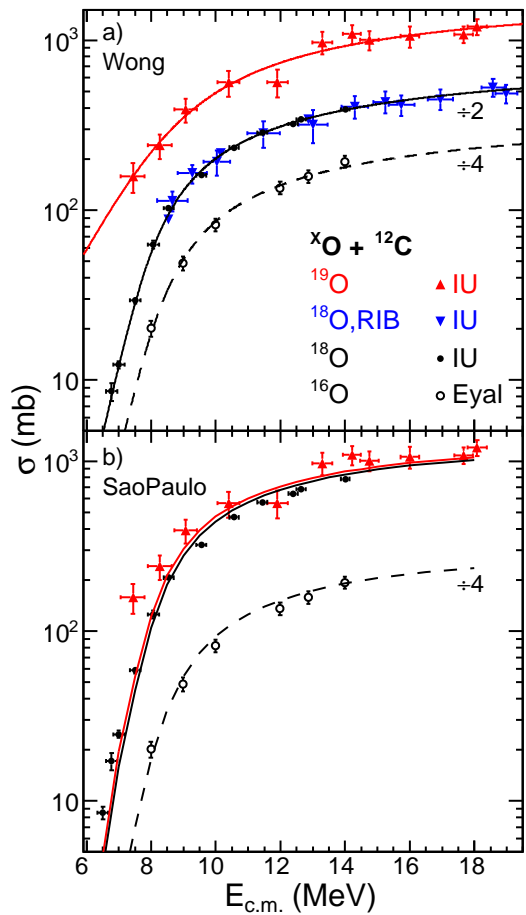}
\caption{\label{fig:xsect} (Color online) 
Fusion excitation function for $^{16,18,19}$O ions incident on $^{12}$C target nuclei. 
Top panel: 
For clarity, the cross-sections for the $^{18}$O and $^{16}$O induced reactions  
have been scaled down by a factor of two and four respectively. 
Lines correspond to fits of experimental data with the Wong formalism. 
Bottom panel: 
Lines correspond to predictions of the RMF+Sao Paulo model. 
While the cross-sections for the $^{18}$O reaction are unscaled, the $^{16}$O 
reaction has been scaled down by a factor of four.
See text for details.}
\end{figure}

Presented in Fig.~\ref{fig:xsect} is the dependence of the fusion cross-section on incident energy for
$^{19}$O + $^{12}$C (up triangles) and $^{18}$O + $^{12}$C (down triangles) measured in the present experiment. Also
shown for comparison is a prior high resolution measurement (closed circles) \cite{Steinbach14a} for 
$^{18}$O + $^{12}$C. 
This high resolution measurement utilized a direct high quality beam of $^{18}$O with a similar 
experimental setup to the present experiment \cite{Steinbach14,Steinbach14a}.
It is clear from Fig.~\ref{fig:xsect} that the measured cross-section for 
the $^{18}$O beam in the present experiment (down triangles) 
is in good agreement with the previous high resolution measurements (closed circles), 
providing confidence in the radioactive beam cross-sections simultaneously measured in the present experiment. 
Also shown in Fig.~\ref{fig:xsect} (open circles) is the fusion excitation function for $^{16}$O + $^{12}$C \cite{Eyal76}. 
The data presented \cite{Eyal76} utilizes only direct 
measurement of evaporation residues to characterize the excitation function. 
This excitation function has also been measured by detection of 
$\gamma$ rays in a thick-target experiment. Although the data from this $\gamma$-ray experiment \cite{Cujec76} is in reasonably 
good agreement with the excitation function depicted, 
the thick target measurements are subject to different uncertainties as compared to the direct detection of ERs. 
Consequently, we omit the thick target data in order to 
make the most straightforward and relevant comparison.

All of the excitation functions depicted in Fig.~\ref{fig:xsect}a manifest the same general trend. 
With decreasing incident energy the cross-section decreases as expected for a barrier controlled process.
Closer examination of the $^{19}$O and $^{18}$O reactions reveals that
the $^{19}$O data exhibits a larger fusion cross-section 
as compared to the $^{18}$O data at essentially all energies measured. 
The most important feature of the measured excitation functions is that at the lowest energies measured the fusion 
cross-section for the $^{19}$O system decreases more gradually with decreasing energy than does the $^{18}$O system.
In order to better quantify these differences in the fusion excitation functions 
we have fit the measured cross-sections with a  
one dimensional barrier penetration model. The Wong formalism \cite{Wong73} considers the 
penetration of an inverted parabolic barrier with the cross-section given by: 
\begin{equation}
\sigma = \frac{R_C^2}{2E}\hbar\omega \cdot ln \left \{ 1+exp\left [\frac{2\pi}{\hbar\omega}(E-V_C)\right] \right \}
\end{equation} 
where $E$ is the incident energy, $V_C$ is the barrier height, $R_C$ is the radius of interaction, and $\hbar\omega$ 
is the barrier curvature. 
The fit of the high resolution $^{18}$O data and 
the $^{16}$O data are indicated as the solid black and dashed black lines in Fig.~\ref{fig:xsect}a respectively. 
The solid red curve in Fig.~\ref{fig:xsect}a depicts the fit of the $^{19}$O data. 
With the exception of the cross-section measured at E$_{\mathrm{c.m.}}$ $\approx$ 12 MeV, 
the measured cross-sections are also reasonably described by the Wong formalism. 
The extracted parameters for the $^{16}$O, $^{18}$O,  and $^{19}$O reactions are summarized in Table 1.
It is not surprising that the barrier height, $V_C$, remains essentially the same for all of the three reactions examined as 
the charge density distribution is essentially unchanged. With increasing neutron number 
an increase in $R_C$ is observed as one
might expect. The most significant change in the Wong parameters is a substantial increase in the magnitude of 
$\hbar\omega$ for the $^{19}$O case
reflecting a narrower barrier, indicating an increased attractive nuclear potential.

\begin{table}[h]
\centering
\caption{\label{tab:5/tc}Wong fit parameters for the indicated fusion excitation functions. See text for details.}
\begin{ruledtabular}
\begin{tabular}{|c|c|c|c|}

 & $V_C$ (MeV) & $R_C$ (fm) & $\hbar$$\omega$ (MeV) \\
\hline
$^{16}$O + $^{12}$C & 7.93 $\pm$ 0.16 & 7.25 $\pm$ 0.25 & 2.95 $\pm$ 0.37 \\

$^{18}$O + $^{12}$C & 7.66 $\pm$ 0.10 & 7.39 $\pm$ 0.11 & 2.90 $\pm$ 0.18\\

$^{19}$O + $^{12}$C & 7.73 $\pm$ 0.72 & 8.10 $\pm$ 0.47 & 6.38 $\pm$ 1.00 \\

\end{tabular}
\end{ruledtabular}

\end{table}

In order to better understand whether the fusion enhancement observed for $^{19}$O is 
associated with changes in the 
nuclear structure 
or an increased role of dynamics as the neutron number is increased , 
we have calculated the fusion
of oxygen isotopes with $^{12}$C nuclei with a simple model. To disentangle the role of structure 
from dynamics, fusion of the two nuclei has been calculated with a Sao Paulo model \cite{Gasques04}. 
As 
the Sao Paulo model does not include any dynamics it
allows one to assess the changes in the fusion cross-section due solely to the changes in the 
density distributions of the nuclei. These density distributions
have been calculated with a relativistic mean field theory \cite{Ring96,Serot86}. As expected, within the RMF theory 
with increasing neutron number the
tail of the neutron density distribution increases slightly while the proton density distribution 
is largely unchanged.
The cross-section predicted from the RMF+Sao Paulo model is depicted in Fig.~\ref{fig:xsect}b. 
While the dashed (black) and solid (black) curves 
provide a reasonable description of the $^{16}$O and $^{18}$O fusion with a $^{12}$C nucleus, the model 
predicts essentially no fusion enhancement
in the case of $^{19}$O relative to $^{18}$O.

\begin{figure}
\includegraphics[scale=0.65]{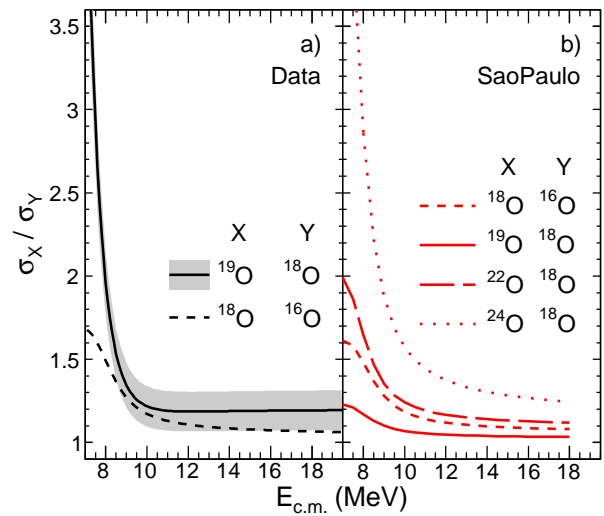}
\caption{\label{fig:xs_ratio} 
Left panel: Dependence of the measured relative fusion cross-section $\sigma$($^{19}$O)/$\sigma$($^{18}$O) and 
$\sigma$($^{18}$O)/$\sigma$($^{16}$O) on E$_{\mathrm{c.m.}}$. The shaded region depicts
the uncertainty associated with the ratio for the $^{19}$O reaction. Right panel: Relative fusion cross-section
predicted for $\sigma$($^{18}$O)/$\sigma$($^{16}$O), $\sigma$($^{19}$O)/$\sigma$($^{18}$O),
$\sigma$($^{22}$O)/$\sigma$($^{18}$O), and $\sigma$($^{24}$O)/$\sigma$($^{18}$O)
by the RMF+Sao Paulo model.
}
\end{figure}

Having parameterized the measured cross-sections using the Wong formalism, it is possible
to more quantitatively examine the observed fusion enhancement and compare the result with the 
RMF+Sao Paulo model.
The dependence of the measured ratios of  $\sigma$($^{19}$O)/$\sigma$($^{18}$O) and 
$\sigma$($^{18}$O)/$\sigma$($^{16}$O) on E$_{\mathrm{c.m.}}$ are shown in Fig.~\ref{fig:xs_ratio}a. 
The solid line corresponds the ratio of the Wong fit for $^{19}$O to the Wong fit for $^{18}$O. 
For E$_{\mathrm{c.m.}}$ $>$ 10 MeV, 
this ratio is essentially constant at a value of approximately 1.2.
In the energy domain above the barrier one expects the ratio of the two cross-sections to be governed by the ratio of 
their geometric cross-sections namely the square of the ratio of their radii.  
The 20\% increase in the cross-section observed for $^{19}$O relative to $^{18}$O can thus be associated with a larger radius for 
$^{19}$O as compared to $^{18}$O.
For E$_{\mathrm{c.m.}}$ $<$ 10 MeV, the quantity $\sigma$($^{19}$O)/$\sigma$($^{18}$O) increases rapidly reaching a value 
of $\approx$ 3 at the lowest energy measured. For reference, it is useful to examine the behavior of 
$\sigma$($^{18}$O)/$\sigma$($^{16}$O) 
presented as the dashed line in Fig.~\ref{fig:xs_ratio}a.
At energies well above the barrier $\sigma$($^{18}$O)/$\sigma$($^{16}$O)  is essentially flat at a value of $\approx$ 1.1.
As one approaches the barrier it increases to a value of approximately 1.7. Hence the 
enhancement observed for $^{19}$O is significantly larger than in the case of $^{18}$O.

The cross-section ratios predicted by the RMF+Sao Paulo model are presented in Fig.~\ref{fig:xs_ratio}b. 
Although the behavior of the ratio $\sigma$($^{18}$O)/$\sigma$($^{16}$O) (dashed red line)  is reasonably good agreement with 
the experiment, for $\sigma$($^{19}$O)/$\sigma$($^{18}$O) (solid red line) the model fails to reproduce the observed trend.
At high energies, E$_{\mathrm{c.m.}}$ $>$ 10 MeV,  
the predicted relative cross-section has a value of $\approx$ 1.03, significantly less 
than the experimentally observed value of $\approx$ 1.2. 
In the near barrier domain, E$_{\mathrm{c.m.}}$ $<$ 10 MeV, 
the underprediction by the theoretical calculation is even more dramatic. The model
fails to exhibit the marked increase with decreasing energy that is experimentally observed,  
only increasing to a value of 1.2 at E$_{\mathrm{c.m.}}$= 7 MeV. 
This failure of the RMF+Sao Paulo model 
to accurately describe the 
change in the fusion cross-section with addition of a neutron to $^{18}$O at both 
high and low energies 
is significant. 
Either the density distributions provided by the RMF model do not accurately represent the tail of 
the neutron density distribution or dynamics unaccounted for in the Sao Paulo model is important. 

To assess whether a modest change in the neutron density distribution increases the relative fusion 
cross-section predicted by the RMF+Sao Paulo model, we also show the results for
$\sigma$($^{22}$O)/$\sigma$($^{18}$O), and 
$\sigma$($^{24}$O)/$\sigma$($^{18}$O) in
Fig.~\ref{fig:xs_ratio}b. 
As the neutron number increases and the tail of the neutron density distribution extends further out,
a larger enhancement of the relative cross-section is observed at near and sub-barrier energies. For
$^{22}$O, and $^{24}$O at E$_{\mathrm{c.m.}}$= 7 MeV an enhancement of 2.0, and 5.0 is predicted.
As it is unreasonable for the tail of the $^{19}$O neutron density distribution to be 
intermediate between
$^{22}$O, and $^{24}$O, we deduce that the observed experimental enhancement indicates the importance of fusion dynamics.
As the two nuclei approach each other at near barrier energies, excitation and polarization of the nuclei can occur,
enhancing the fusion process. Unpaired neutrons, due to their weaker binding, may be 
particularly effective in enhancing fusion. Systematic measurement of fusion excitation functions for 
isotopic chains of neutron-rich light nuclei would provide a robust means for elucidating both the 
influence of nuclear structure as well as dynamics on the fusion cross-section.
While the present comparison with the static RMF+Sao Paulo model represents a first step in 
demonstrating that dynamics plays an important role in the fusion of neutron-rich light nuclei, 
comparison with more sophisticated models that directly incorporate dynamics is necessary.
Efforts to compare the present experimental data with a density constrained time dependent 
Hartree-Fock model \cite{Keser12} are presently underway. 

In summary, we have developed an experimental technique to measure fusion induced by low-intensity 
radioactive beams (10$^3$-10$^6$ ions/s). 
Using this approach we have
measured for the first time the fusion of $^{19}$O + $^{12}$C at incident 
energies near and below the barrier to investigate whether fusion of neutron-rich light nuclei is 
enhanced relative to their $\beta$ 
stable isotopes and may serve as a heat source triggering X-ray superbursts in an accreting neutron star.
Comparison of the fusion excitation function for $^{19}$O + $^{12}$C with that of $^{18}$O + $^{12}$C, 
clearly demonstrates that for the $^{19}$O system, fusion is significantly 
enhanced. Well above the barrier this enhancement is approximately 20 \% which can be related 
to an increase in the 
radius due just to the presence of the additional neutron. Near and below the barrier the 
fusion enhancement 
is even more dramatic, increasing to a factor of three at the lowest energy measured. 
To ascertain if the fusion enhancement can be understood simply in terms of the extended tail of 
the neutron density distribution, we have compared the experimental results with a static fusion model.
Using proton and neutron density distributions calculated with 
a relativistic mean field (RMF) model, the Sao Paulo fusion model
fails to predict the measured enhancement. Only for extremely 
neutron-rich nuclei does the RMF+Sao Paulo model predict an enhancement comparable in magnitude to 
the measured enhancement, suggesting the importance of dynamics 
in the fusion of neutron-rich light nuclei.
This result motivates measurement of fusion, using this technique, for even 
more neutron-rich oxygen nuclei at energies near and below the barrier to elucidate the 
defining characteristics of this fusion enhancement \footnote{An experiment to measure fusion in $^{39,47}$K + $^{28}$Si
at MSU-NSCL is presently underway and an experiment to measure fusion in 
$^{20,21}$O + $^{12}$C at GANIL has been proposed.}.

We wish to acknowledge the support of the staff at Florida State University's John D. Fox accelerator in 
providing the high quality beam that 
made this experiment possible. Development of the Sao Paulo model code by Dr. Helber Dussan is gratefully acknowledged.
This work was supported by the U.S. Department of Energy 
under Grant No. DE-FG02-88ER-40404 (Indiana University), DE-FG02-87ER40365 (Indiana University Nuclear Theory), 
DE-SC0008808 (NUCLEI SciDAC Collaboration),
Grant no. DE-FG02-02ER-41220 (Florida State University) and
the National Science Foundation under Grant No PHY-1491574 (Florida State University). J.V. acknowledges the 
support of a NSF Graduate Research Fellowship
under Grant No. 1342962.

\bibliography{fusion_19O}

\begin{thebibliography}{27}
\expandafter\ifx\csname natexlab\endcsname\relax\def\natexlab#1{#1}\fi
\expandafter\ifx\csname bibnamefont\endcsname\relax
  \def\bibnamefont#1{#1}\fi
\expandafter\ifx\csname bibfnamefont\endcsname\relax
  \def\bibfnamefont#1{#1}\fi
\expandafter\ifx\csname citenamefont\endcsname\relax
  \def\citenamefont#1{#1}\fi
\expandafter\ifx\csname url\endcsname\relax
  \def\url#1{\texttt{#1}}\fi
\expandafter\ifx\csname urlprefix\endcsname\relax\def\urlprefix{URL }\fi
\providecommand{\bibinfo}[2]{#2}
\providecommand{\eprint}[2][]{\url{#2}}

\bibitem[{\citenamefont{Burbridge et~al.}(1957)}]{Burbridge57}
\bibinfo{author}{\bibfnamefont{E.~M.} \bibnamefont{Burbridge}}
  \bibnamefont{et~al.}, \bibinfo{journal}{Rev. Mod. Phys.}
  \textbf{\bibinfo{volume}{29}}, \bibinfo{pages}{547} (\bibinfo{year}{1957}).

\bibitem[{\citenamefont{Goriely et~al.}(2011)}]{Goriely11}
\bibinfo{author}{\bibfnamefont{S.}~\bibnamefont{Goriely}} \bibnamefont{et~al.},
  \bibinfo{journal}{Astro. Phys. Jour. Lett.} \textbf{\bibinfo{volume}{738}},
  \bibinfo{pages}{L32} (\bibinfo{year}{2011}).

\bibitem[{\citenamefont{Khuyagbaatar et~al.}(2014)}]{Khuyagbaatar14}
\bibinfo{author}{\bibfnamefont{J.}~\bibnamefont{Khuyagbaatar}}
  \bibnamefont{et~al.}, \bibinfo{journal}{Phys. Rev. Lett}
  \textbf{\bibinfo{volume}{112}}, \bibinfo{pages}{172501}
  (\bibinfo{year}{2014}).

\bibitem[{\citenamefont{Yanez et~al.}(2014)}]{Yanez14}
\bibinfo{author}{\bibfnamefont{R.}~\bibnamefont{Yanez}} \bibnamefont{et~al.},
  \bibinfo{journal}{Phys. Rev. Lett.} \textbf{\bibinfo{volume}{112}},
  \bibinfo{pages}{152702} (\bibinfo{year}{2014}).

\bibitem[{\citenamefont{Oliphant et~al.}(1935)\citenamefont{Oliphant, Kempton,
  and Rutherford}}]{Oliphant35}
\bibinfo{author}{\bibfnamefont{M.~L.~E.} \bibnamefont{Oliphant}},
  \bibinfo{author}{\bibfnamefont{A.~E.} \bibnamefont{Kempton}},
  \bibnamefont{and}
  \bibinfo{author}{\bibfnamefont{L.}~\bibnamefont{Rutherford}},
  \bibinfo{journal}{Proc. Royal Soc. A} \textbf{\bibinfo{volume}{150}},
  \bibinfo{pages}{241} (\bibinfo{year}{1935}).

\bibitem[{\citenamefont{Horowitz et~al.}(2008)\citenamefont{Horowitz, Dussan,
  and Berry}}]{Horowitz08}
\bibinfo{author}{\bibfnamefont{C.~J.} \bibnamefont{Horowitz}},
  \bibinfo{author}{\bibfnamefont{H.}~\bibnamefont{Dussan}}, \bibnamefont{and}
  \bibinfo{author}{\bibfnamefont{D.~K.} \bibnamefont{Berry}},
  \bibinfo{journal}{Phys. Rev. C} \textbf{\bibinfo{volume}{77}},
  \bibinfo{pages}{045807} (\bibinfo{year}{2008}).

\bibitem[{\citenamefont{Cumming and Bildsten}(2001)}]{Cumming01}
\bibinfo{author}{\bibfnamefont{A.}~\bibnamefont{Cumming}} \bibnamefont{and}
  \bibinfo{author}{\bibfnamefont{L.}~\bibnamefont{Bildsten}},
  \bibinfo{journal}{The Astrophys. Jou.} \textbf{\bibinfo{volume}{559}},
  \bibinfo{pages}{L127} (\bibinfo{year}{2001}).

\bibitem[{\citenamefont{Liang et~al.}(2003)}]{Liang03}
\bibinfo{author}{\bibfnamefont{J.~F.} \bibnamefont{Liang}}
  \bibnamefont{et~al.}, \bibinfo{journal}{Phys. Rev. Lett.}
  \textbf{\bibinfo{volume}{91}}, \bibinfo{pages}{152701}
  (\bibinfo{year}{2003}).

\bibitem[{\citenamefont{Loveland et~al.}(2006)}]{Loveland06}
\bibinfo{author}{\bibfnamefont{W.}~\bibnamefont{Loveland}}
  \bibnamefont{et~al.}, \bibinfo{journal}{Phys. Rev. C}
  \textbf{\bibinfo{volume}{74}}, \bibinfo{pages}{064609}
  (\bibinfo{year}{2006}).

\bibitem[{\citenamefont{Lemasson et~al.}(2009)}]{Lemasson09}
\bibinfo{author}{\bibfnamefont{A.}~\bibnamefont{Lemasson}}
  \bibnamefont{et~al.}, \bibinfo{journal}{Phys. Rev. Lett}
  \textbf{\bibinfo{volume}{103}}, \bibinfo{pages}{232701}
  (\bibinfo{year}{2009}).

\bibitem[{\citenamefont{Kolata et~al.}(2012)}]{Kolata12}
\bibinfo{author}{\bibfnamefont{J.~J.} \bibnamefont{Kolata}}
  \bibnamefont{et~al.}, \bibinfo{journal}{Phys. Rev. C}
  \textbf{\bibinfo{volume}{85}}, \bibinfo{pages}{054603}
  (\bibinfo{year}{2012}).

\bibitem[{\citenamefont{Kohley et~al.}(2013)}]{Kohley13}
\bibinfo{author}{\bibfnamefont{Z.}~\bibnamefont{Kohley}} \bibnamefont{et~al.},
  \bibinfo{journal}{Phys. Rev. C} \textbf{\bibinfo{volume}{87}},
  \bibinfo{pages}{064612} (\bibinfo{year}{2013}).

\bibitem[{\citenamefont{Wu and Bertsch}(1986)}]{Wu86}
\bibinfo{author}{\bibfnamefont{J.~Q.} \bibnamefont{Wu}} \bibnamefont{and}
  \bibinfo{author}{\bibfnamefont{G.~F.} \bibnamefont{Bertsch}},
  \bibinfo{journal}{Nucl. Phys. A} \textbf{\bibinfo{volume}{457}},
  \bibinfo{pages}{401} (\bibinfo{year}{1986}).

\bibitem[{\citenamefont{Rudolph et~al.}(2012)}]{Rudolph12}
\bibinfo{author}{\bibfnamefont{M.~J.} \bibnamefont{Rudolph}}
  \bibnamefont{et~al.}, \bibinfo{journal}{Phys. Rev. C}
  \textbf{\bibinfo{volume}{85}}, \bibinfo{pages}{024605 (R)}
  (\bibinfo{year}{2012}).

\bibitem[{\citenamefont{Carnelli et~al.}(2014)}]{Carnelli14}
\bibinfo{author}{\bibfnamefont{P.~F.~F.} \bibnamefont{Carnelli}}
  \bibnamefont{et~al.}, \bibinfo{journal}{Phys. Rev. Lett.}
  \textbf{\bibinfo{volume}{112}}, \bibinfo{pages}{192701}
  (\bibinfo{year}{2014}).

\bibitem[{\citenamefont{Wiedenh{\"o}ver et~al.}(2012)}]{RESOLUT}
\bibinfo{author}{\bibfnamefont{I.}~\bibnamefont{Wiedenh{\"o}ver}}
  \bibnamefont{et~al.}, in \emph{\bibinfo{booktitle}{Fifth International
  Conference on Fission and Properties of Neutron-rich Nuclei}}, edited by
  \bibinfo{editor}{\bibfnamefont{J.}~\bibnamefont{Hamilton}} \bibnamefont{and}
  \bibinfo{editor}{\bibfnamefont{A.}~\bibnamefont{Ramayya}}
  (\bibinfo{publisher}{World Scientific}, \bibinfo{year}{2012}), ISBN
  \bibinfo{isbn}{978-981-4525-42-8}.

\bibitem[{\citenamefont{deSouza et~al.}(2011)}]{deSouza11}
\bibinfo{author}{\bibfnamefont{R.~T.} \bibnamefont{deSouza}}
  \bibnamefont{et~al.}, \bibinfo{journal}{Nucl. Instr. and Meth.}
  \textbf{\bibinfo{volume}{A632}}, \bibinfo{pages}{133} (\bibinfo{year}{2011}).

\bibitem[{\citenamefont{Nicolis and R.Beene}(1993)}]{evapor}
\bibinfo{author}{\bibfnamefont{N.~G.} \bibnamefont{Nicolis}} \bibnamefont{and}
  \bibinfo{author}{\bibfnamefont{J.}~\bibnamefont{R.Beene}},
  \bibinfo{journal}{unpublished}  (\bibinfo{year}{1993}).

\bibitem[{\citenamefont{Steinbach et~al.}(2014{\natexlab{a}})}]{Steinbach14a}
\bibinfo{author}{\bibfnamefont{T.~K.} \bibnamefont{Steinbach}}
  \bibnamefont{et~al.}, \bibinfo{journal}{Phys. Rev. C.}
  \textbf{\bibinfo{volume}{90}}, \bibinfo{pages}{041603(R)}
  (\bibinfo{year}{2014}{\natexlab{a}}).

\bibitem[{\citenamefont{Steinbach et~al.}(2014{\natexlab{b}})}]{Steinbach14}
\bibinfo{author}{\bibfnamefont{T.~K.} \bibnamefont{Steinbach}}
  \bibnamefont{et~al.}, \bibinfo{journal}{Nucl. Instr. and Meth.}
  \textbf{\bibinfo{volume}{A743}}, \bibinfo{pages}{5}
  (\bibinfo{year}{2014}{\natexlab{b}}).

\bibitem[{\citenamefont{Eyal et~al.}(1976)\citenamefont{Eyal, Beckerman,
  Checkhik, Fraenkel, and Stocker}}]{Eyal76}
\bibinfo{author}{\bibfnamefont{Y.}~\bibnamefont{Eyal}},
  \bibinfo{author}{\bibfnamefont{M.}~\bibnamefont{Beckerman}},
  \bibinfo{author}{\bibfnamefont{R.}~\bibnamefont{Checkhik}},
  \bibinfo{author}{\bibfnamefont{Z.}~\bibnamefont{Fraenkel}}, \bibnamefont{and}
  \bibinfo{author}{\bibfnamefont{H.}~\bibnamefont{Stocker}},
  \bibinfo{journal}{Phys. Rev. C} \textbf{\bibinfo{volume}{13}},
  \bibinfo{pages}{1527} (\bibinfo{year}{1976}).

\bibitem[{\citenamefont{Cujec and Barnes}(1976)}]{Cujec76}
\bibinfo{author}{\bibfnamefont{B.}~\bibnamefont{Cujec}} \bibnamefont{and}
  \bibinfo{author}{\bibfnamefont{C.~A.} \bibnamefont{Barnes}},
  \bibinfo{journal}{Nucl. Phys. A} \textbf{\bibinfo{volume}{266}},
  \bibinfo{pages}{461} (\bibinfo{year}{1976}).

\bibitem[{\citenamefont{Wong}(1973)}]{Wong73}
\bibinfo{author}{\bibfnamefont{C.~Y.} \bibnamefont{Wong}},
  \bibinfo{journal}{Phys. Rev. Lett.} \textbf{\bibinfo{volume}{31}},
  \bibinfo{pages}{766} (\bibinfo{year}{1973}).

\bibitem[{\citenamefont{Gasques et~al.}(2004)}]{Gasques04}
\bibinfo{author}{\bibfnamefont{L.~R.} \bibnamefont{Gasques}}
  \bibnamefont{et~al.}, \bibinfo{journal}{Phys. Rev. C}
  \textbf{\bibinfo{volume}{69}}, \bibinfo{pages}{034603}
  (\bibinfo{year}{2004}).

\bibitem[{\citenamefont{Ring}(1996)}]{Ring96}
\bibinfo{author}{\bibfnamefont{P.}~\bibnamefont{Ring}}, \bibinfo{journal}{Prog.
  Part. Nucl. Phys.} \textbf{\bibinfo{volume}{37}}, \bibinfo{pages}{193}
  (\bibinfo{year}{1996}).

\bibitem[{\citenamefont{Serot and Walecka}(1986)}]{Serot86}
\bibinfo{author}{\bibfnamefont{B.~D.} \bibnamefont{Serot}} \bibnamefont{and}
  \bibinfo{author}{\bibfnamefont{J.~D.} \bibnamefont{Walecka}},
  \bibinfo{journal}{Adv. Nucl. Phys.} \textbf{\bibinfo{volume}{16}},
  \bibinfo{pages}{1} (\bibinfo{year}{1986}).

\bibitem[{\citenamefont{Keser et~al.}(2012)\citenamefont{Keser, Umar, and
  E.Oberacker}}]{Keser12}
\bibinfo{author}{\bibfnamefont{R.}~\bibnamefont{Keser}},
  \bibinfo{author}{\bibfnamefont{A.~S.} \bibnamefont{Umar}}, \bibnamefont{and}
  \bibinfo{author}{\bibfnamefont{V.}~\bibnamefont{E.Oberacker}},
  \bibinfo{journal}{Phys. Rev. C} \textbf{\bibinfo{volume}{85}},
  \bibinfo{pages}{044606} (\bibinfo{year}{2012}).

\end{thebibliography}


\end{document}